\documentclass[twocolumn]{aa}
\usepackage{graphicx}
\usepackage{txfonts}
\begin{document}
   \title{Evidence for new physics from  clusters ?}

   \author{Alain Blanchard
\inst{1}
          \and
           Marian Douspis
\inst{2}
          }

   \offprints{Alain Blanchard}

   \institute{LATT, 14 avenue Edouard Belin, F-31400 Toulouse, France\\
              \email{alain.blanchard@ast.obs-mip.fr}
           \and
	      LATT, 14 avenue Edouard Belin, F-31400 Toulouse, France\\
               \email{douspis@ast.obs-mip.fr}
             }
\date{}

   \abstract{The abundance of local clusters is a traditional way to
   derive the amplitude of matter fluctuations, commonly specified by
   $\sigma_8$, but which suffers from a systematic uncertainty arising
   from the lack of accurate knowledge of the mass temperature
   relation.  In the present work, by assuming that the observed
   baryon content of clusters is representative of the universe, we
   show that the mass temperature relation ($M-T$) can be specified
   for any cosmological model.  WMAP constraints on the baryonic
   content of the Universe and the $\Omega_M-H_0$ relation allows one
   further improvement in tightening this $M-T$ relation.  This
   approach allows to remove most of the above uncertainty, and to
   provide an estimation of $\sigma_8$ whose uncertainty is
   essentially statistical.  The values we obtain are fortuitously
   almost independent of the matter density of the Universe
   ($\sigma_8\sim 0.6-0.63$) with an accuracy better than 5\%.  Quite
   remarkably, the amplitude of matter fluctuations can be also
   tightly constrained to similar accuracy from existing CMB
   measurements alone, once the dark matter content is specified.
   However, the amplitude inferred in this way in a concordance model
   ($\Lambda-CDM$) is significantly larger than the value derived from
   the above method based on X-ray clusters.  Such a discrepancy would
   almost disappear if the actual optical thickness of the Universe
   was 0 but could also be alleviated from more exotic solutions: for
   instance the existence of a new non-baryonic light dark component
   in the Universe as massive neutrinos, with $\Omega_d \sim
   0.01-0.03$.  However, recent other indications of $\sigma_8$ favor
   a high normalization. In this case, the assumption that the
   baryonic content observed in clusters actually reflects the
   primordial value has to be relaxed : either there exists a large
   baryonic dark component in the Universe with $\Omega_d \sim
   0.01-0.03 \sim 0.5 \Omega_{b}$ or baryons in clusters have
   undergone a large depletion during the formation of these
   structures.  We concluded that the baryon fraction in clusters is
   not representative and therefore that an essential piece of the
   physics of baryons in clusters is missing in standard structure
   formation scenario.

 }

   \maketitle
%

\section{Introduction}

 The amplitude of matter fluctuations in the present-day universe is
 an important quantity of cosmological relevance. The abundance of
 clusters is an efficient way to evaluate this quantity, commonly
 expressed by $\sigma_8$, the {\em r.m.s.} amplitude of the matter
 fluctuations on the $8 h^{-1}\rm Mpc$ scale. A statistical precision
 of a few \% on $\sigma_8$ is possible from existing samples of X-ray
 clusters, but in practice the relation between mass and temperature
 is needed for such evaluation:
\begin{equation}
T =A_{TM} M^{2/3}_{15}(\Omega_M (1+\Delta)/179)^{1/3}h^{2/3}(1+z)\rm \;\; keV
\label{eq:tm}  
\end{equation}
  (Oukbir and Blanchard, 1992), $\Omega_M$ being the present--day
  matter density parameter and $\Delta$ being the contrast density
  relative to the Universe at the radius at which $M_{15}$ is
  taken. The value of $A_{TM}$ has been estimated from X-ray
  properties of clusters by different methods, essentially hydrostatic
  equations on one side and numerical simulations on the other side,
  which lead to  different normalizations (from $\sigma_8
  \sim 0.6$ to $\sigma_8 \sim 1.$). Mass lensing measurements of
  clusters could in principle provide a direct measurement of this
  quantity but present-day results are contradictory. This question
  remained unresolved because the amplitude of matter obtained from
  clusters with hydrostatic equations leads to low values, $\sigma_8
  \sim 0.7 \pm 0.06$ (Markevitch 1998, Reiprich et al. 2002, Seljak
  2002) while WMAP recently obtained $\sigma_8 \sim 0.9 \pm 0.1$
  (Spergel et al.  2003).  However, the virial cluster masses are
  difficult to obtain, and values inferred with different methods
  spread a large range of values; for instance Roussel et al. (2000)
  pointed out that hydrostatic mass estimations were lower than values
  inferred from numerical simulations and Henry (2004) recently found
  that published values of $\beta$, a quantity proportional to
  $A_{TM}$, could differ by a factor of nearly two. This leaves a
  large uncertainty on the actual amplitude of matter fluctuations
  derived from clusters. In this paper, we propose a new approach to
  derive the mass-temperature relation in a self-consistent way. This
  allows us to combine the baryon budget from the CMB, observed gas mass
  in clusters and the present day abundance of clusters to infer a
  tight constraint on the amplitude of matter fluctuations obtained
  from the cluster abundance and compare it to the amplitude inferred
  from the CMB.

\section{Mass-Temperature relation}

\subsection{The $\sigma_8$-$A_{TM}$ degeneracy.}

\begin{figure}
   \centering 
	\includegraphics[width=8cm]{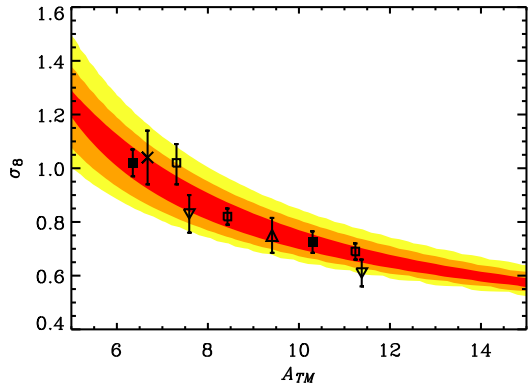} 
	\caption{The central
   area is  the amplitude of matter fluctuations
   expressed in term of $\sigma_8$ versus the normalization of the
   mass temperature $A_{\rm TM}$ (Eq.~\ref{eq:tm}) for a flat model
   with $\Omega_M = 0.3$. Grey areas are our one, two and three sigma
   level contours. Filled squares are from Vauclair et al. (2003),
   open squares are Pierpaoli et al. (2001), around $A_{\rm TM}\sim
   7$ and Pierpaoli et al. (2003), $\times$ symbol is from Evrard et
   al. (2002), Triangle is from Seljak (2002), inverted triangles are
   from Viana et al. (2002) and Viana et al. (2003).}  
	\label{Sigatm}
\end{figure}

 The determination of $\sigma_8$ from the cluster abundance is a
 standard procedure that has been used by many authors, leading to
 somewhat dispersed values.  Here we use the Sheth and Tormen (1999)
 mass function and a sample of X-ray selected local clusters ($f_x
 \leq 2.2 10^{-11}$ erg/s/cm$^2$ and $|b| \leq 20 \deg$, Blanchard et
 al., 2000 { updated from BAX, Sadat et al. 2004}). The relation
 between $\sigma_8$-$A_{TM}$ is presented in Figure \ref{Sigatm} for a
 flat model with $\Omega_M = 0.3$ with some other recent measurements,
 based on ROSAT samples of X-ray clusters and a recent analytical mass
 functions (del Popolo 2004).  We do not include analyses based on
 HEAO-1 such as Henry (2004) or using the classical Press and
 Schechter expression as do Ikebe et al. (2001), or other mass
 functions.  A NFW profile (Navarro et al. 1995) with $c = 5$ was
 assumed when necessary. This shows that most of the dispersion among
 different analyses (which use nearly the same clusters) is due to the
 different values used for the normalization constant $A_{TM}$.  Most
 of the remaining differences are due to differences in temperatures
 used (with or without cooling flow correction, or temperature
 cuts). Notice that the point presenting the largest deviation (Viana
 et al. (2002)) is based on the luminosity function. We converted
 their mean luminosity to a mean temperature of 2.68 keV to derive an
 equivalent $A_{TM}$.

\subsection{The baryon fraction argument}

\begin{figure}
   \centering
  \includegraphics[width=8cm]{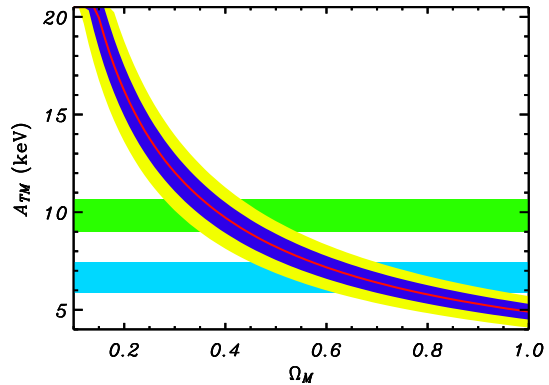}
      \caption{The red line is the central value of 
$A_{\rm TM}$ which is the normalization of the mass temperature
relation Eq.~\ref{eq:tm}. The WMAP relation between $H_0$ and
$\Omega_0$ has been used, as well as the constraint on the baryon
content of the Universe. One and two $\sigma$ uncertainties arising
from uncertainty on $\Omega_{b}$ ($= 0.023\pm 0.002$)  are shown as blue and yellow
areas. Horizontal areas correspond to estimations of $A_{\rm TM}$ from
hydrostatic methods (light green)  obtained by Roussel et al. (2000) and 
Markevitch (1998) and from numerical simulations
(light blue)  obtained from Bryan \& Norman (1998) and  Evrard, Metzler \&  Navarro (1996).}
         \label{atm}
   \end{figure}

 Clusters are useful cosmological probes in several important
 ways. Their baryonic fraction $f_b$ can be inferred from
 observations:
\begin{equation}
f_b =  \frac{M_b}{M_{tot}} ;
\label{eq:fb}
\end{equation}
 the mass in observed baryons, $M_b$, consists mainly of the X-ray gas
 and of a small part of the stars (Roussel et al 2000), while the
 total mass $M_{tot}$ could be estimated through one of the
 above--mentioned methods. Under the assumption that the baryonic and
 dark matter amounts are representative of the universe, the baryon
 fraction can be related to the cosmological parameters density
 $\Omega_{b}$ and $\Omega_{m}$:
\begin{equation}
f_b =  \Upsilon \frac{\Omega_{b}}{\Omega_{m}} .
\label{eq:fbb}
\end{equation}
 $\Upsilon$ is a numerical factor that has to be introduced in order
 to correct for the depletion of gas during cluster formation and
 which can be determined only from numerical simulations (White et al.
 1993).  In practice, a good working value, at least in the outer part
 of clusters, is $\Upsilon = 0.925$ (Frenk et al. 1999).  The baryonic
 content of the Universe is now known quite accurately through WMAP
 and other CMB measurements ($\omega_{b}= \Omega_{b}h^2 = 0.023\pm
 0.002$, Spergel et al. 2003; the statistical uncertainty being
 doubled in order to account for differences in various priors),
 essentially consistent with the abundance of Deuterium (Kirkman et
 al. 2003) and with the baryonic content of the IGM (Tytler et al
 2004).

\subsection{Self-consistent mass-temperature relation}

 While the above relations have been widely used to obtain constraints
 on $\Omega_m$ assuming that the $M-T$ relation is known, here we
 follow a different approach. Given the uncertainty in the actual
 value of $A_{TM}$, we can use the knowledge of gas (and stars) masses
 in clusters and of $\Omega_{b}$ to constrain the total mass in
 clusters, at least as a function of $\Omega_m$ and $h$, and thereby
 infer the mass-temperature relation. A slight source of complexity in
 gas mass measurements comes from the fact that the gas in clusters
 may be clumpy. If so, the gas mass estimation from average  radial 
 profile of the emissivity overestimates the actual gas mass by a
 factor $C^{1/2}$, where $C$ is a measurement of the clumping of the
 gas.  Sadat and Blanchard (2001) have studied in detail the change in
 shape of the gas fraction with radius in clusters: they found that
 the gas fraction follows rather well a scaling law, i.e. it is
 roughly identical among different clusters when expressed in term of
 the radius normalized to the virial radius. Furthermore they found
 that in the outer part the shape was close to what has been found in
 numerical simulations provided the outer amplitude is corrected for
 clumping (the value of $\Upsilon$ being roughtly constant $\sim
 0.925$ for $\Delta \leq 1000$). This implies that correction for
 clumping is indeed important to have an unbiased description of the
 internal structure of clusters. Mathiesen et al. (1999) found an
 average $C^{1/2}$ of 1.16 at the radius corresponding to a contrast
 density $\Delta$ of 500 (they also found that taking only clusters
 with no secondary peak at a level of 1\% of the global maximum
 reduced the average $C^{1/2}$ to 1.093).  Because the clumping factor
 seems to vary rapidly with radius, it is safe to work on clusters at
 a similar radius. We have used the gas mass determination from
 Vikhlinin, Forman and Jones (1999, VFJ99 hereafter), using their most
 external radius for mass determination, which is $h-$dependent, at
 the average temperature of 4 keV. VFJ99 provided gas mass
 measurements at the radius R1000 where the contrast density in the
 gas is 1000 times some fiducial baryon density (2.85 $10^9
 M_\odot$/Mpc$^3$), which corresponds to nearly half of the best
 $\omega_{b}$ derived from WMAP.  Typical density contrasts at our
 working radius are in the range 480--625, at which we can directly
 apply the above correction for clumping. VFJ99 excluded clusters with
 double or very irregular X-ray morphology, a criteria that seems less
 demanding than the criteria for regularity used by Mathiesen et
 al. (1999). However, in both cases roughly one third of the clusters
 were excluded from the analysis. We therefore used the value
 $C^{1/2}$ = 1.093.  The difference to $C^{1/2}$ = 1.16 is a source of
 systematic uncertainty on the final mass of 6\%.  We further
 corrected for a star contribution of $34h^{1.5}$\% (Roussel et
 al. 2000).  Knowing the baryon mass, relations~2 and ~3 can be used
 to infer the total mass (depending on $\Omega_m$) { at the radius
 R1000. Finally, in order to use the above mass estimation in the mass
 function we need to estimate the mass at the virial radius
 (unfortunately measurements of both apparent gas mass and clumping
 are not available at the virial radius).} This virial mass can be
 estimated assuming a NFW profile with a fixed concentration parameter
 $c$.  Hereafter we used $c = 5$.  WMAP has provided high precision
 data in a field where order of magnitude estimations were the only
 possibility a few years ago. This allows us to constrain very tightly
 some quantities which are often combinations of a few cosmological
 parameters.  For the present analysis, we use the location of the
 so--called Doppler peak which allows us to establish a tight relation
 between $\Omega_m$ and $h$ in a flat universe (Page et al. 2003).
 The above procedure has been applied to derive the mass temperature
 normalization $A_{\rm TM}$ as a function of $\Omega_m$. The result is
 shown in Figure 1.  As one can see the above procedure allows us to
 determine the value of $A_{\rm TM}$ as a function of $\Omega_m$ with
 a small uncertainty : we found $A_{\rm TM} \sim 4.9\Omega_m^{-0.75}
 \pm 10\%$, for $0.3 \leq \Omega_m \leq 1.0$; the values cover the
 range of the various estimations based on the different approaches
 that we have mentioned.  From this relation we can now infer the
 typical temperature of clusters formed from fluctuations within $R =
 8h^{-1}$ Mpc spheres:
\begin{equation}
T_{8h^{-1}{\rm Mpc}} \approx A_{TM} (1.19 \Omega_m)^2/3 \approx 3.65 \rm keV \Omega_m^{-0.09} 
\end{equation} 
 therefore, the amplitude of matter fluctuations $\sigma_8$ is
 essentially controlled by the abundance of clusters with temperatures
 around 3.5 keV almost {\em independently of the value $\Omega_m $ }
 (this comes from the fact that $A_{TM}$ varies with $\Omega_m $ and
 that this variation accidentally compensates almost exactly the
 variation obtained for a fixed value of $A_{TM}$).  Knowing the
 mass-temperature relation and its uncertainty we can determine the
 amplitude of matter fluctuations { by fitting the local temperature
 distribution function and assuming a $\Gamma$-like spectrum with
 $\Gamma = 0.2$} as explained in section 2.1.

 The result is shown in Figure 2a.  As one can see, at a given value
 of $\Omega_M$ the amplitude of $\sigma_8$ is well constrained.
 Furthermore to the first order the best $\sigma_8$ is independent of
 $\Omega_M$ ($\sigma_8 \sim 0.63 \pm 3.\% (1 \sigma)$ for
 $\Omega_\Lambda = 0.7$. Interestingly this is close to the value
 obtained by Viana et al. (2002): $\sigma_8 \sim 0.61$). Our
 conclusion appears somewhat surprising as it differs from standard
 analyses based on a fixed normalization $A_{TM}$, which cannot
 simultaneously account for the baryon fraction in a consistent way
 for arbitrary $\Omega_M$.  Gas masses from the VFJ99 sample present a
 moderate dispersion of the order of 20\% (Sadat et al. 2005),
 implying rather small uncertainties on our gas fraction estimates of
 the order of about 4\% at our working radii, which will produce an
 uncertainty on $\sigma_8$ of 2.5\%. More important is the correction
 for clumping.  For instance, Voevodkin \& Vikhlinin (2004) have
 estimated $\sigma_8$ from the baryon mass function in a Cold Dark
 Matter framework. In the case $\Omega_M \sim 0.3$ their approach is
 very close to ours, but they used gas mass estimation at the virial
 radius and did not correct for clumping.  This leads to virial masses
 which are $\sim 20\%$ lower than ours, leading to $A_{TM} \sim 11$ keV and
 therefore $\sigma_8 \sim 0.7$, in very good agreement with their
 estimation.

   \begin{figure}[!t]
   \centering
   \includegraphics[width=8cm]{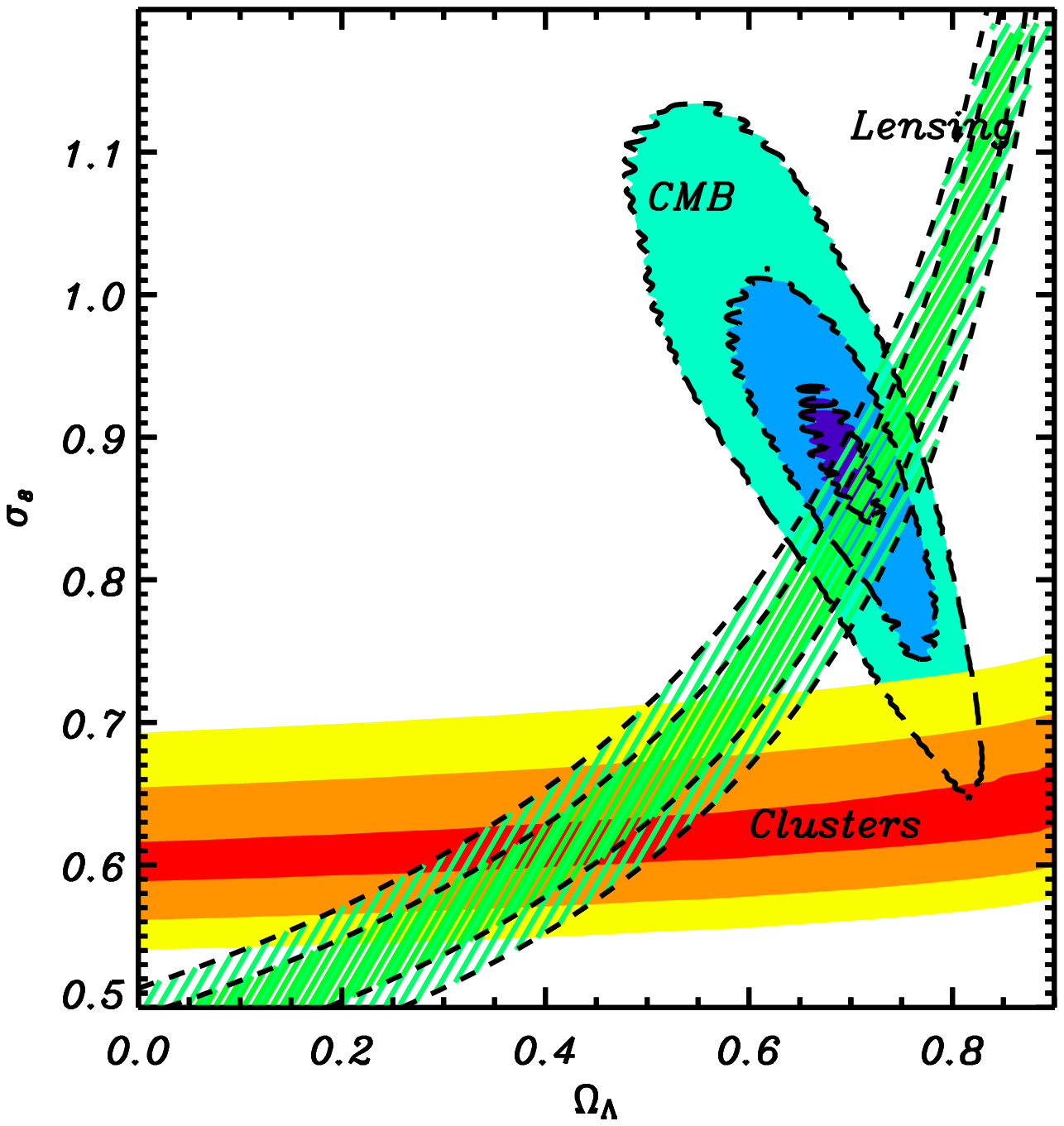}
   \includegraphics[width=8cm]{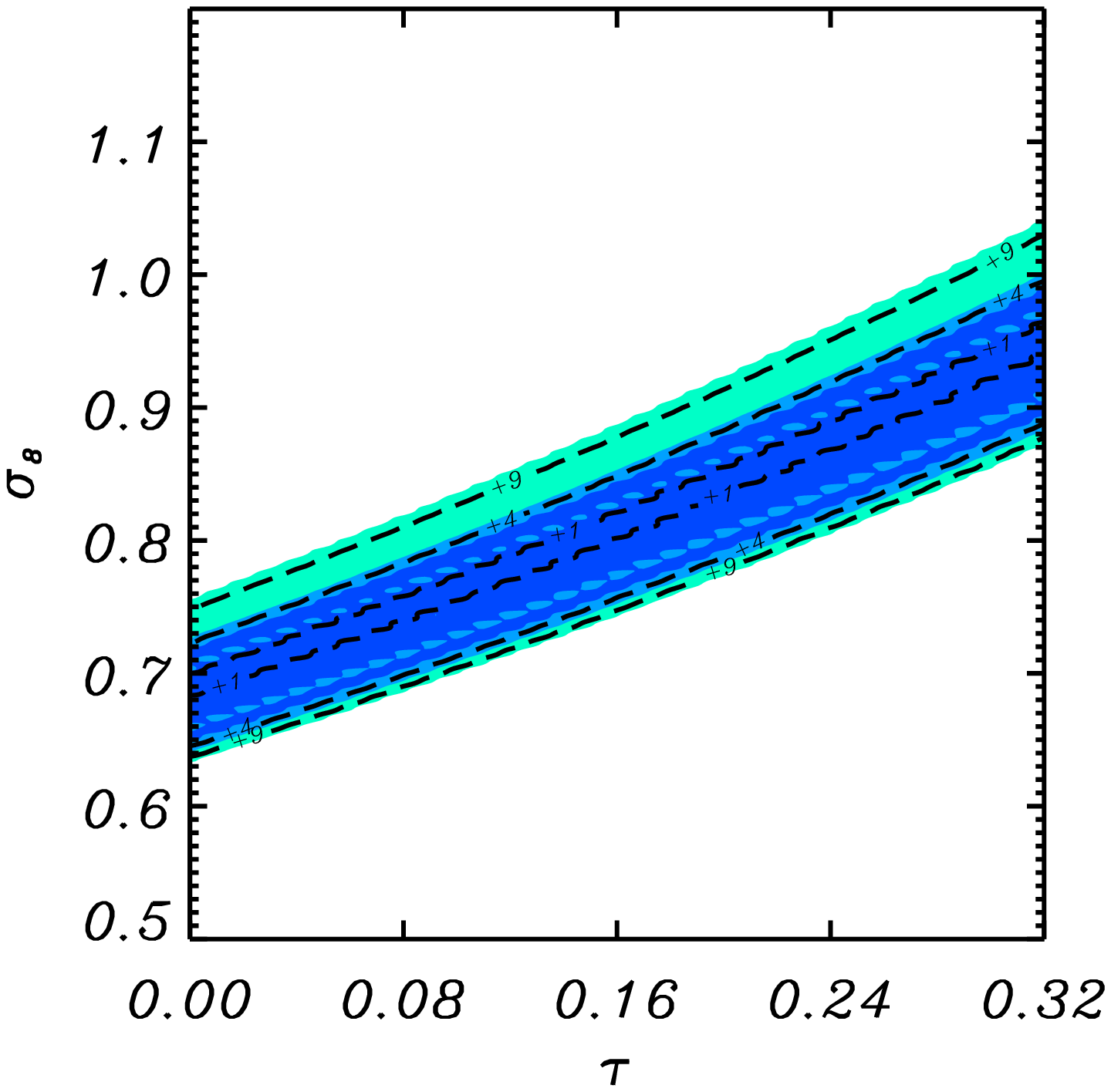}
      \caption{a) top: The amplitude of matter fluctuations from clusters 
abundance using the mass--temperature relation found in the present
analysis compared to the amplitude of matter fluctuations derived from
CMB data (Grainge et al 2003, Pearson et al. 2003, Ruhl et al. 2003,
Kuo et al, 2004). The grey area corresponds to 1, 2, 3 $\sigma$ contours
on two parameters, dashed lines are contours on one parameter. The one and two 
sigma  amplitudes obtained from an average of recent weak shear 
measurements are also shown as dashed regions
(see text for references). b) bottom: amplitude   
of matter fluctuations from CMB data versus optical thickness.            }
         \label{sl}
   \end{figure}

\section{Need for Dark Matter}

 The amplitude of matter fluctuations is
 strongly constrained by the CMB data. In the following we use the
 constraint on $\sigma_8$ in a concordance model obtained from the CMB
 fluctuation analysis including the temperature--polarization cross
 power spectrum (TE) by the WMAP team (Kogut et al 2003).

 The comparison of the value of $\sigma_8 $ from CMB data with the one
 from clusters reveales a critical discrepancy between the two
 measurements (Figure 2a).  It is clear that within any model with
 $\Omega_\Lambda \sim 0.7$ the amplitude of $\sigma_8$ we derived from
 clusters, $\sigma_8 = 0.63 \pm 0.02$, is significantly smaller than
 what is expected from the CMB alone ($\sigma_8 = 0.88 \pm 0.035$),
 which is close to the recent determination from the combination of
 WMAP and the Lyman $\alpha$ forest (Seljak et al 2004).

 The non-zero optical thickness $\tau$, which is requested only from
 the TE spectrum, is a key factor in this discrepancy: the high
 $\sigma_8 $ obtained from CMB data depends critically on the actual
 amplitude of the optical thickness $\tau$ (see Figure 2b), although
 forcing $\tau = 0$ does not entirely remove the discrepancy.  An
 accurate knowledge of $\tau$ is therefore critical to properly
 evaluate the amplitude of matter fluctuations in the concordance
 model. One can see from Figure 2b how much the value of $\sigma_8 $
 obtained from CMB data depends on the actual value of the optical
 thickness and remains the main source of uncertainty in establishing
 the value of $\sigma_8 $. We have also checked that when CMB data are
 restricted to the range $400\leq l \leq 1200$, the above discrepancy
 remains essentially unchanged.  Allowing a non-power law initial
 power spectrum is therefore not expected to solve this issue.

 Here above, we have considered models in which the dark matter is
 only made of cold dark matter, the dark energy being a pure
 cosmological constant (in terms of the equation of state of vacuum $p
 = w \rho$, this means $w = -1$), and that X-ray gas and known stars
 are the only existing baryons in clusters. A first possibility to
 investigate is to examine whether a different equation of state for
 the vacuum, so-called quintessence, might solve this discrepancy.  We
 have therefore investigated flat models with arbitrary $w$ and
 quintessence content $\Omega_Q$. Indeed combinations of CMB and
 cluster data are known to provide tight constraints on such models
 (Douspis et al. 2003).  With the approach developed here, models
 which were found to match CMB and clusters were found to satisfy the
 following constraints: $0.46 <\Omega_Q < 0.54$ and $ - 0.5 < w < -
 0.4 $.  Such models are currently at odds with constraints on
 quintessential models (Douspis et al.  2003; Tegmark et al. 2003;
 Riess et al. 2004) resulting from the combination of various data
 including type Ia supernovae data. We therefore require an
 alternative approach to solve the above issue.  In the following, we
 examine whether the introduction of an additional component of the
 dark matter content of the universe would remove the above
 discrepancy. Neutrinos are known to exist and to be massive, so
 perhaps the most natural massive component of the universe to be
 introduced is in the form of a neutrino contribution.  This solution
 has already been advocated to solve this discrepancy in an Einstein
 de Sitter Universe (Elgar{\o}y \& Lahav 2003, Blanchard et
 al. 2003). Indeed, the presence of a light, but non-zero, component
 of dark matter significantly modifies the transfer function of
 primordial fluctuations which results in a lower amplitude on small
 scales. Given existing measurements of mass differences we consider
 only the case where the masses are equal. Within a concordance model
 ($\Omega_\Lambda = 0.7$ $\Omega_m = 0.3$), by combining the
 constraints from CMB and cluster data, and marginalizing on
 ($\omega_{\rm b}$, $H_0$, $n$, $\tau$) we found that a contribution
 of $\Omega_\nu = 0.016 \pm 0.003$ is preferred with a significance
 level well above 3$\sigma$ (see Figure 3a),  improving the
 significance of such possible evidence compared to Allen et
 al. (2003).  This confirms that the presence of a small contribution
 of neutrinos, with a typical mass of .25 eV, to the density of the
 universe allows one to reconcile the amplitude of matter fluctuations
 from clusters with the one inferred from CMB data. We notice that
 such a value is above the upper limit inferred by the WMAP team using
 a combination of several astronomical data (Spergel et
 al. 2003). Finally, weak shear estimations have provided measurements
 of the amplitude of matter fluctuations which can be compared to that
 obtained from clusters (Refregier 2003). There are some differences
 in published values which probably reflect systematic uncertainties
 not yet fully identified. However, taking the independent
 measurements of $\sigma_8$ from weak lensing obtained from an average
 of recent measurements (Bacon et al. 2003, Brown et al. 2004, Chang
 et al. 2004, Hamana et al. 2003, Heymans et al. 2004, Hoekstra et
 al. 2002, Jarvis et al. 2003, Massey et al., 2004, Refregier et
 al. 2002, Rhodes et al. 2004, Van Waerbeke et al., 2004) which lead
 to an acceptable $\chi^2$, from WMAP and Lyman $\alpha$ forest
 (Seljak et al 2004), and the value of $\beta$ from 2dFGRS (Hawkins et
 al. 2003), the low amplitude of $\sigma_8$ obtained above is not
 favored. We are therefore left with the conclusion that our initial
 assumption that baryons in clusters are fairly representative of
 baryons in the universe is unlikely and therefore that the observed
 amount of baryons in clusters does not reflect the actual primordial
 value (a possibility that has been advocated by Ettori 2003). Several
 mechanisms could lead to this situation: the most direct way could be
 the fact that a significant fraction of the baryons are in a dark
 form, either in the Universe or in clusters (for instance either in
 the form of Machos, or in a large gaseous unidentified component,
 Bonamente et al, 2003), or that a significant fraction of the baryons
 has been expelled from clusters during their formation process.  In
 such cases, the observed $M_b$ is biased to lower values.  The actual
 mass of clusters from Eq.~2 and Eq.~3 can then be obtained assuming a
 depletion factor $1-f$ implying that $f\Omega_b$ represents the
 missing baryons.  Again the combination of CMB and cluster
 constraints allows us to evaluate the amplitude of $f\Omega_b$.  From
 Figure 3b, one can see that such a component, $f\Omega_b \sim 0.023$,
 should represent nearly half ($\Upsilon \sim 0.5$) of the primordial
 baryons in order to solve the discrepancy. Although heating processes
 are required to account for the observed properties of X-ray
 clusters, they currently do not lead to such a high level of
 depletion (Bialek, Evrard \& Mohr 2001).

   \begin{figure}
   \centering
   \includegraphics[width=8cm]{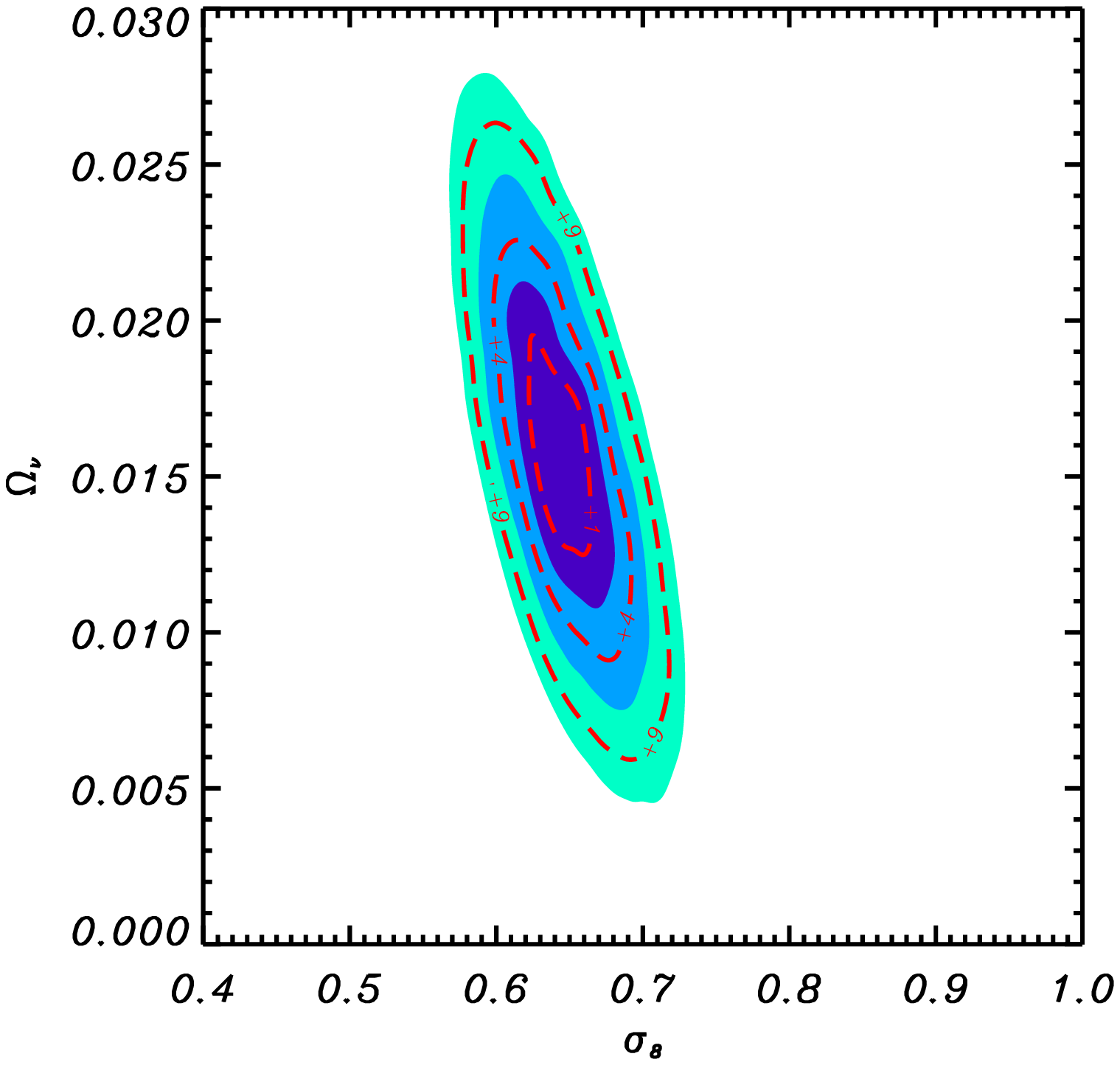}
   \includegraphics[width=8cm]{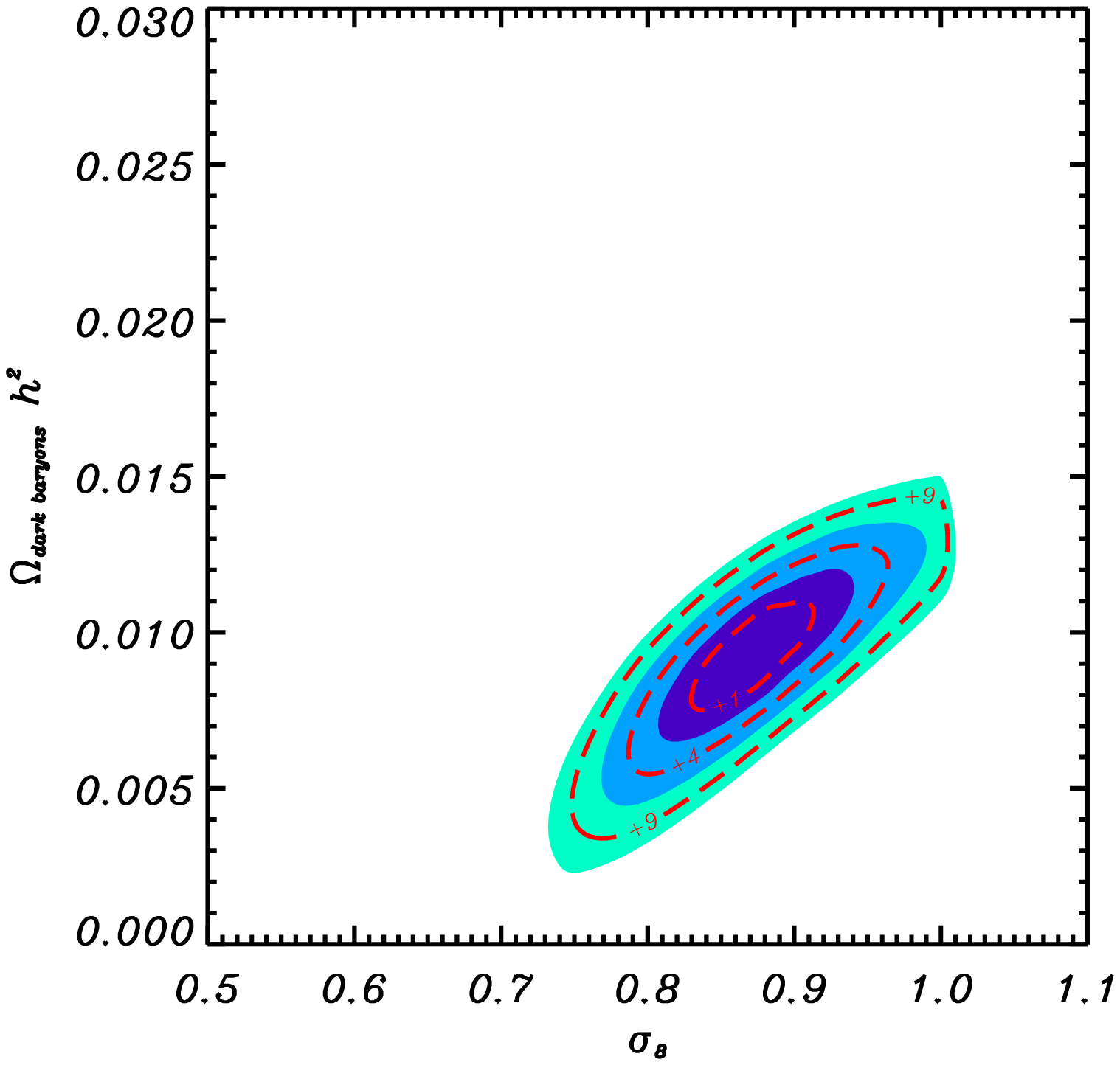}

      \caption{Constraints on $\Omega_\nu$ and $\Omega_{dark baryon}h^2$ 
    given by the combined analysis of CMB and Cluster data. The dark energy 
component  $\Omega_\Lambda $ has been set to $0.7$ in a flat cosmology.           }
         \label{powplot3}
   \end{figure}

\section{Conclusions}

 The determination of the amplitude of matter fluctuations within pure
 cold dark matter model, using two methods, namely the CMB and the
 local cluster abundance, leads to two  significantly
 different values.  There are several ways  to solve  this
 discrepancy, although each represents a noticeable departure
 from the standard concordance model.  The existence of a non-baryonic
 dark component, like a neutrino contribution, would allow us to solve
 this discrepancy, although such a solution leads to a low value of
 $\sigma_8$ which is not favored by  other evidence.  If the
 actual value is larger, $\sigma_8
\sim 0.8-0.9$, the unavoidable conclusion is that the baryonic content
of clusters {\bf at $\Delta \sim 500$ } is not representative of the
Universe.  In this case, an astrophysical solution could be that
baryons in clusters could be in a dark form, or at least undetected
until now. Alternatively, baryons in clusters could have been severely
depleted implying that the actual value $\Upsilon$ is much smaller
than the value we used above, the apparent baryon fraction being
biased to low values compared to the actual primordial value.  Finally, several
observations might help to clarify this issue: the above conclusion
relies on the actual value of the optical depth $\tau$ found by
WMAP. If the actual value was consistent with zero most of the
discrepancy would disappear. Confirmation of the actual value of $\tau$
is therefore critical and its better estimation will allow a better
estimation of $\sigma_8$ from the CMB. Other sources of information on
$\sigma_8$ will also obviously clarify this issue: weak lensing
can potentially allow one to directly measure  the actual amplitude of
matter fluctuations with a similar precision to what has been obtained
here with clusters, provided that systematic uncertainties are fully
understood; the clusters masses could be measured from their lensing
signal providing a direct estimation of the normalization constant
$A_{TM}$, allowing one to distinguish between low and high
normalizations. Other direct measurements of the amplitude of matter
fluctuations like those derived from the Lyman--$\alpha$ forest power
spectrum (Croft et al. 1998) could also help clarify  this issue. It
is remarkable that some of the observations that are expected in the
near future can potentially bring fundamental information on
clusters physics { or alternatively} may reveal the existence of a
previously unidentified type of dark matter with $\Omega_{DM}$ as low
as 0.01.

\begin{acknowledgements}
      
 This research has made use of the X-ray cluster database (BAX)
which is operated by the Laboratoire d'Astrophysique de Toulouse-Tarbes (LATT),
under contract with the Centre National d'Etudes Spatiales (CNES). We
acknowledge useful comments from the referee which contributed  improving
the  paper.
\end{acknowledgements}

\end{document}